\documentclass{IEEEcsmag}

\usepackage[colorlinks, urlcolor=blue, linkcolor=blue, citecolor=blue]{hyperref}

\usepackage{upmath}

\jvol{XX}
\jnum{XX}
\paper{8}
\jmonth{May/June}
\jname{IT Professional}
\pubyear{2021}

\usepackage{multirow}
\usepackage{lineno}
\usepackage[table]{xcolor}

\begin{document}
    \sptitle{Human-AI Interaction for Graphics and Visualization}

    \title{Design Exploration of AI-assisted Personal Affective Physicalization}

    \author{Ruishan Wu}
    \affil{Simon Fraser University, Burnaby, BC, V5A 4Y7, Canada}

    \author{Zhuoyang Li}
    \affil{Eindhoven University of Technology, Eindhoven, North Brabant, 5612 AZ, The Netherlands}

    \author{Charles Perin}
    \affil{University of Victoria, Victoria, BC, V8P 5C2, Canada}

    \author{Sheelagh Carpendale}
    \affil{Simon Fraser University, Burnaby, BC, V5A 4Y7, Canada}

    \author{Can Liu}
    \affil{City University of Hong Kong, Hong Kong, 523808, China}

    \markboth{Human-AI Interaction for Graphics and Visualization}{Human-AI Interaction for Graphics and Visualization}

    \begin{abstract}
        \looseness-1
        Personal Affective Physicalization is the process by which individuals express emotions through tangible forms to record, reflect on, and communicate. Yet such physical data representations can be challenging to design due to the abstract nature of emotions. Given the shown potential of AI in detecting emotion and assisting design, we explore opportunities in AI-assisted design of personal affective physicalization using a Research-through-Design method. 
        We developed PhEmotion, a tool for embedding LLM-extracted emotion values from human-AI conversations into parametric design of physical artifacts. A lab study was conducted with 14 participants creating these artifacts based on their personal emotions, with and without AI support. We observed nuances and variations in participants' creative strategies, meaning-making processes and their perceptions of AI support in this context. We found key tensions in AI-human co-creation that provide a nuanced agenda for future research in AI-assisted personal affective physicalization.
    \end{abstract}

    \maketitle


\chapteri{D}ata physicalization has been shown as an effective method to support personal reflection on self-tracking data~\cite{Thudt-self}, including emotional, behavioral and health data. Existing emotion self-tracking methods primarily let people create free-form objects through low-barrier crafting activities, such as forming shapes with clay~\cite{lee2017designing}. While these crafting activities can serve the purpose of making expressive objects, not everyone is good at handcrafting thus may not have sustained interest in engaging with them. One potential way to engage people in such activities is to use computational methods to help make the physicalizations beautiful, interesting objects that they would like to carry, wear or place in their living environment.

The advancement of Generative AI (Gen-AI) brought power to conversational user interfaces (CUIs), which have been increasingly used for emotional support and analysis~\cite{li2025human}. Mental Health support chatbots, such as platforms like Replika\footnote{\url{https://replika.com/}}, Woebot\footnote{\url{https://woebothealth.com/}}, and Wysa\footnote{\url{https://www.wysa.com/}}, as well as general-purpose assistants (e.g., ChatGPT, Gemini), are used by millions for comfort, stress reduction, and for practicing coping skills~\cite{li2025human}. Gen-AI-powered CUIs are effective in eliciting emotional expressions about memories and feelings, thus could be used as an engagement method in the process of self-tracking of emotion data~\cite{li2025human}. 
We explore questions around whether the analytical and generative power of today's Gen-AI models can be leveraged to create an engaging data physicalization process. Ideally, this process will make it easier for people to create physicalizations that they find aesthetically pleasing and emotionally meaningful. To our knowledge, this question has not been explored before.

In this work, we describe an experimental exploration to shed light on possible ways to use AI to bridge emotional expression and physicalization design. 
We use parametric design to bridge computation and crafting, by translating human-AI conversations into emotional values that are then mapped to shape parameters for affective physicalizations. This is realized by developing a research probe --- PhEmotion, a tool that combines a CUI for eliciting, analyzing and quantifying emotions, and an interface for users to map quantified emotions to shape parameters of 3D-printable physical artifacts.
Through an in-lab observation study, we observed participants using PhEmotion to create metaphorical, ambiguous, and emotionally resonant personal affective physicalizations. We documented their perceptions of Gen AI support in this context, and how and why they interact with Gen AI in their design process.

We present this research not as a demonstration of AI superiority in design, but as a provocation to increase discussion around these questions: 
a) How do people create a physical embodiment of complex emotions?,
b) Does Human-AI co-creation benefit personal affective physicalization? and
c) Is there an interplay between automation vs. manual control?

Our work makes two primary contributions:
\begin{itemize}
    \item The design and implementation of a research probe - PhEmotion, supporting an AI-assisted workflow for designing affective physicalization.
    \item Empirical findings that highlight key tensions in AI-human co-creation of affective physicalization: i) the tension between emotional autonomy and ownership, and algorithmic guidance;
ii) the tension between people's need to navigate emotional design through ambiguous and abstract metaphors, and precise metrics; 
and iii) the tension between the benefits of AI assistance and the need for human-driven meaning-making and personal narrative. 
\end{itemize}

\section{Background}

\begin{figure}
    \centering
    \includegraphics[width=\linewidth]{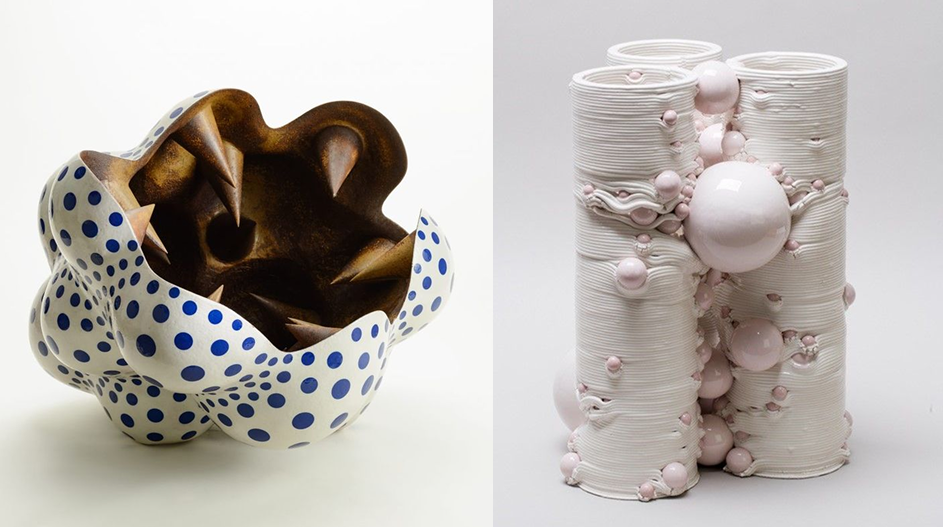}
    \caption{(left) Harumi Nakashima uses glaze textures, clay shapes, and wave-point decorations to express anxiety and dread. (right) Andrea Salvatori combines ceramic 3D printing with manual techniques, breaking and deforming regular patterns to reflect feelings of rebellion and freedom.}
    \label{fig:artwork}
\end{figure}

\subsection{Personal Affective Data Physicalization}

Data physicalization is defined as the practice of representing data through tangible objects. Formally, a data physicalization is described as "a physical artifact whose geometry or material properties encode data"~\cite{jansen2015opportunities}. Prior research has demonstrated cognitive and emotional benefits of physical data representations, notably within personal informatics and self-reflection contexts \cite{Thudt-self, houben2016physikit}.

When applied to personal affective data—information about emotions, moods, or sentiments—physicalization introduces unique opportunities and challenges due to the subjective and nuanced nature of emotional experiences. Emotional data is inherently qualitative, often lacking explicit, universally agreed-upon metrics. This positions personal affective physicalization distinctly apart from traditional scientific visualizations that prioritize objective legibility and precise quantification. Instead, in our study, affective physicalization aligns more closely with personal, reflective data visualizations that explore the intersection of quantitative and qualitative boundaries.

Recent literature highlights two primary interpretations of "affective visualization": one that emphasizes accurately representing emotional data, such as mood fluctuations or sentiment analysis~\cite{wang2023emotional}, and another that focuses on using visual or physical encodings specifically to evoke emotional responses from viewers~\cite{lan2023affective, wang_emotional_2019}.

In this paper, we define "personal affective physicalization" as an integrative concept merging both of these perspectives. Primarily, our research involves physicalizing emotional data—transforming personal emotional states into tangible forms to support self-reflection. Simultaneously, we acknowledge that physical artifacts inherently possess expressive qualities that can evoke or communicate emotional meanings through choices in material, texture, and form ~\cite{feng2022tactile}. Artists frequently utilize tactile properties in their works to evoke emotions, suggesting that aesthetic and tactile encodings significantly influence emotional perception (see Figure~\ref{fig:artwork}). Our approach thus intentionally bridges representational accuracy with affective expressiveness, aligning with both analytical and evocative interpretations from prior literature.

\subsection{Parametric Design for Physicalization}

Creating data physicalizations typically involves translating numerical data into physical properties, a process commonly termed ``rendering'' data physically~\cite{djavaherpour2021}. 
One method for achieving this is \emph{parametric design}—a computational technique where digital models are algorithmically generated through adjustable parameters (e.g., shape curvature, size, texture density). Changing input parameters dynamically produces various forms, effectively linking data to physical geometries~\cite{daneshzand2023kiriphys}. Parametric design tools, such as Grasshopper®\footnote{\url{https://www.grasshopper3d.com}}, provide powerful functionality, but require domain-specific expertise, which can limit accessibility for non-experts interested in creating personal or emotionally meaningful data physicalizations.

Existing parametric approaches primarily cater to structured, quantitative datasets, easily mapped onto physical dimensions~\cite{djavaherpour2021}. In contrast, emotional data often lacks clear numerical representation due to its subjective complexity. Therefore, traditional parametric pipelines designed for numerical clarity may inadequately handle nuanced emotional information. Recognizing this gap motivates our research to integrate parametric design with accessible interfaces, aiming to enable lay users to translate subjective emotional states into personalized physical artifacts.

\subsection{Constructive and Narrative Physicalization Approaches}

Two conceptual frameworks—constructive and narrative physicalizations—further inform our research approach. \emph{Constructive physicalization} emphasizes interactive, user-driven exploration, enabling individuals to physically build or modify data representations through tangible components, facilitating direct engagement and experimentation~\cite{huron_constructive_2014}. \emph{Narrative physicalization} focuses on storytelling and emotional communication, embedding data within curated, symbolic objects designed to deliver specific insights or evoke emotional responses from the audience~\cite{karyda_narrative_2021}.

We combine aspects of both frameworks: employing constructive elements by providing users interactive agency in artifact creation, while also leveraging narrative elements to enable emotional reflection through personalized storytelling.

\subsection{AI and Large Language Models for Personal Affective Physicalization}

Recent advancements in generative Artificial Intelligence (AI) and Large Language Models (LLMs), notably OpenAI's GPT series\footnote{\url{https://chatgpt.com/}}, present significant opportunities to enhance personal affective physicalization processes~\cite{basole2024generative, park2025reimagining}. 
From the framework in personal data physicalization identified by Thudt et al.~\cite{Thudt-self}, people have to prepare and summarize relying on the contextual knowledge to design a data summary, by logging their experiences and preparing tokens from their separated log. There is a whole bunch of work that relies on data analysis and data literacy.
LLMs can process natural language descriptions, summarizing data attributes from abstract emotional inputs. This capability may lower barriers for non-expert users by translating descriptive emotional narratives into data design suggestions, potentially enriching the expressiveness and accessibility of personal affective physicalizations.

Nevertheless, integrating AI into physicalization design necessitates careful attention to human-AI collaboration dynamics. Balancing automated generation and human creative autonomy is crucial to ensure emotional authenticity and user agency, avoiding overly generic or prescriptive outputs~\cite{basole2024generative}. Offering AI suggestions and user-editable design recommendations form essential elements of our approach, as we aim to maintain user agency and creative control.

    \section{PhEmotion - Design and Implementation}

We adopted a Research-through-Design (RtD) approach \cite{zimmerman2007research} to design and implement PhEmotion as a research probe, which provides us the opportunity to explore \emph{how individuals perceive, use, and interpret AI automation in the context of emotional expression through physical artifacts}.

\begin{figure*}[h!]
    \centering
    \includegraphics[width=1.0\linewidth]{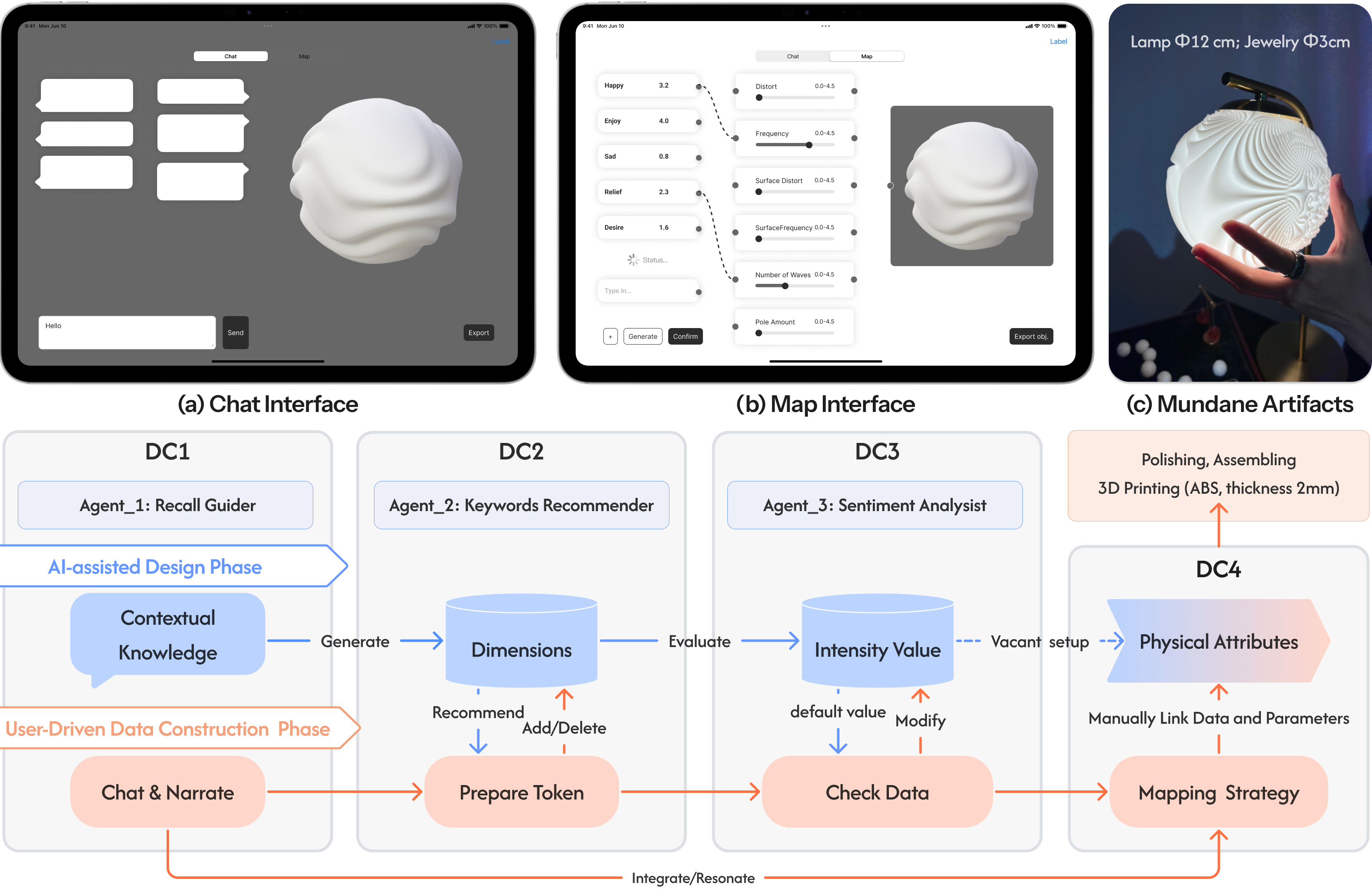}
    \caption{
   Workflow diagram of the PhEmotion system. Steps involving AI/LLM assistance are marked in blue, while actions relying primarily on human input or decision-making are marked in pink. PhEmotion employs AI assistance across design components DC1--DC3 in the design phase, collecting emotional context and leveraging semantic analysis to suggest affective tokens and associated intensity values. At the same time, human autonomy is preserved throughout DC1--DC4 in the data construction phase, enabling users to freely input emotional narratives, select or adjust affective tokens, modify intensity values, and manually map emotion-value pairs onto physical shape parameters (in DC4, b). The final artifacts can then be exported and 3D printed. Specifically: (a) shows the chat interface corresponding to DC1; (b) illustrates the data-mapping interface covering DC2, DC3, and DC4; and (c) presents sample printed artifacts as outcomes of the workflow.}
    \label{interaction}
\end{figure*}

\subsection{Two–Phase Framework for Personal Affective Physicalization}

We draw on Thudt et al.~\cite{Thudt-self}’s two-phase model (design → data construction) as a theoretical grounding. In our affective context, the first phase emphasizes structuring raw emotional narratives into tokens and intensities, which may appear as a form of ‘data work.’ However, consistent with Thudt et al., we treat these activities as design decisions---determining what to track and how to encode it---rather than as a separate data phase. The subsequent construction phase then parallels Thudt et al.'s iterative integration of tokens into evolving physicalizations.

The \textbf{design phase} involves creating a \emph{data summary} based on \emph{contextual knowledge} and developing \emph{physical mappings} between data attributes and material properties. During this phase, users determine \emph{what} data to track, \emph{how} to summarize it through appropriate scales and ranges, and \emph{how} to encode it physically through mapping design decisions. The \textbf{data construction phase} encompasses the iterative process of \textit{logging experiences}, \textit{preparing tokens}, and integrating them into the evolving \textit{physicalization}. In the data construction phase, users repeatedly document new experiences, create physical tokens representing these data points, and incorporate them into a growing physical artifact. These two phases are interconnected through realization (design → construction) and adjustments (construction → design), enabling continuous refinement and adaptation of both the physical representation and the underlying data structure.

Within PhEmotion, AI support is purposefully concentrated in the \textbf{design phase}, where it assists users in articulating, structuring, and mapping their affective criteria, while the subsequent \textbf{data-construction phase} remains a human-led practice of iterative logging and token integration. This division confronts the central difficulty that felt emotion is a subjective, interpretive phenomenon rather than an objective measurement—self-reports are shaped by context, memory, and social-desirability biases. Accordingly, the system translates these qualitative impressions into quantifiable representations while acknowledging the porous boundary between overtly stated feelings and the tacit undercurrents that accompany them.

Below, we introduce the four major design components (DCs) of PhEmotion, each aligned with its AI-supported yet user-driven approach.

\subsection{Design Components and Rationales}

Figure~\ref{interaction} shows AI-assisted steps in blue and human-led steps in pink.  AI is purposefully concentrated in DC1--DC3 to surface and structure affective data (example prompts could be found in the supplemental materials); DC4 remains human-controlled to study how people appropriate that data in the tangible domain. 

\paragraph{DC1 – Conversational Emotional Elicitation (AI \& Human)}  
\textbf{Goal:} Help users articulate rich, situated emotional narratives.  
\textbf{AI affordance:} We built a prompt scaffold grounded in the “open-ended reflection” micro-skills from motivational interviewing.  Pilot sessions ($n{=}4$) refined the prompt to balance empathetic listening (\textit{“What sights, sounds, or bodily sensations do you recall?”}) with gentle probes that encourage specificity.  
\textbf{Human agency.} Users type freely; the LLM only paraphrases or asks follow-ups if the user pauses for $\ge5$ s. This preserves ownership of the story.

\paragraph{DC2 – Affective Token Extraction (AI $\leftrightarrow$ Human)}  
\textbf{Goal:} Transform raw narrative into a concise but expressive “affective palette.”  
\textbf{AI process:} The LLM applies semantic role labelling and lexicon matching to propose 7 emotion tokens (e.g., \textit{wistful}, \textit{proud}).  
\textbf{Rationale:} Prior work on affective visualization shows that limiting the palette helps novices reason about mappings~\cite{karyda_narrative_2021}.  
\textbf{Human agency:} Users can delete, rename, or add tokens; every edit is logged for later analysis of agency.

\paragraph{DC3 - Intensity Scaling (AI $\leftrightarrow$ Human)}  
\textbf{Goal:} Assign a numeric intensity (0–4.5) to each token.    
\textbf{Rationale:} Providing a \emph{first guess} lowers the barrier to quantification yet invites reflection.
\textbf{Physicalization link:} These values later drive parametric ranges (e.g., curvature radius, surface roughness).

\paragraph{DC4 – Token–Parameter Mapping (Human $\rightarrow$ Parametric Engine)}  
\textbf{Goal:} Connect affective data to geometric attributes.  
\textbf{Rationale:} Mapping is the creative crux where meaning emerges~\cite{Thudt-self}.  Automating it would foreclose the very tension (metric vs.\ metaphor) we seek to study.  
\textbf{Support:} A Grasshopper panel displays live previews; users drag “emotion pins” onto parameter sliders and can observe the 3-D form update in real time.  
\textbf{Outcome:} The finalized mapping matrix $\langle$token,\,intensity,\,parameter$\rangle$ is exported with the OBJ file for fabrication and further reflection.

\begin{figure*}[t!]
    \centering
    \includegraphics[width=0.85\textwidth]{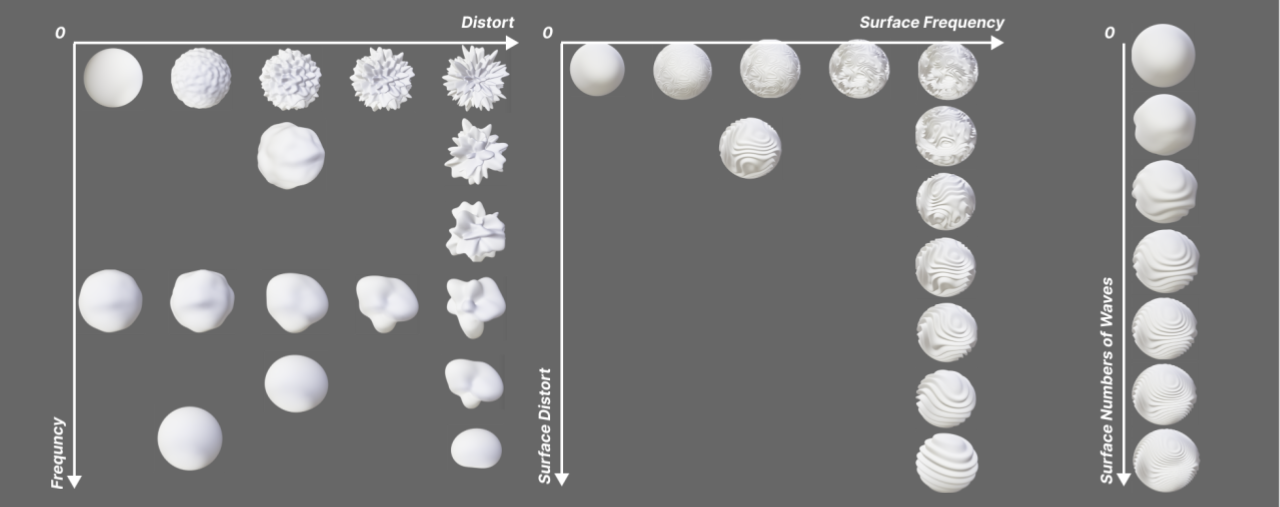}
    \caption{Attribute legend rendered in 3D. The two-dimensional coordinates represent combinations of grouped attributes.}
    \label{legend}
\end{figure*}

\subsection{Shape Parameter (Variables)}
For the purpose of this research, the number of shape parameters (variables) that emotions can be mapped to is relatively small, yet includes a breadth of shape parameter types.
PhEmotion supports both \emph{countable} and \emph{uncountable} shape parameters. 
For example, the number of waves is discrete, while surface distortion and frequency introduce continuous, ambiguous variations that may better map with complex emotional states.
\begin{itemize}
    \item Countable: number of waves.
    \item Uncountable: global distortion and global frequency noise, surface distortion, surface frequency noise.
\end{itemize}

Previous research suggests that when designing physicalizations, people often want to map data values to high-level, dependent variables such as the look and feel, the expansion and the stretchability of a physicalization, rather than low-level, independent variables like the position of a mark or the length of a bar~\cite{daneshzand2023kiriphys, daneshzand2025designing}. 
Therefore, PhEmotion supports shape parameters that operate at a relatively high-level, and that we group into two conceptual categories (see Figure~\ref{legend}):
\begin{itemize}
    \item Surface Texture category: surface distortion, frequency noise, number of waves.
    \item Overall Shape category: global distortion and shape frequency.
\end{itemize}

These two categories are designed to facilitate both clear mapping of emotional states and intuitive understanding of how different emotions might alter the shape. Surface texture attributes are intended to evoke subtle, tactile changes that contribute to the overall feel of the shape, while overall shape attributes manipulate the form on a grander scale, influencing its primary structure and overall impression.

The rationale behind the choice of these shape and specific parameters stems from several factors. Firstly, previous work in the field of tactile and emotional design (e.g., Feng et al. \cite{feng2022tactile}) has predominantly focused on sphere and hemispherical shapes. This serves as a foundation for our design, as these shapes are familiar, simple, and easy to manipulate. Additionally, spherical forms offer a smooth, continuous surface, which provides a suitable canvas for testing how different shape attributes influence emotional perception. We chose to begin with a sphere because variations in its surface are easily noticeable and allow for a clear investigation of emotional mapping.

Building on the foundational sphere attributes from an existing NFT website \footnote{\url{https://blobmixer.14islands.com/}}, we aimed to investigate the shape attributes that have been less thoroughly explored in the current body of research. 
Previous data physicalization studies have often examined the effects of individual attributes, such as size, roughness, spikiness, roundness, and differences between spheres and cylinders \cite{du2024rough, feng2022tactile, jansen2015psychophysical}. By focusing on sphere-based noise attributes—such as surface distortion, frequency, and wave count—we aim to fill a gap in the literature and investigate how these shape attributes can represent complex emotional states more effectively. 
Specifically, we are interested in understanding how the interaction among multiple shape attributes, i.e., the combined effect of altering both surface texture and overall shape parameters, can influence people’s emotional perceptions while designing physicalization. 
This interaction is particularly important because it may reveal how the synergy of multiple parameters can create a more cohesive or expressive emotional experience.

In summary, the selection and setup of these shape attributes in PhEmotion explore the interactive and emotional potential of using multiple shape attributes on a sphere as emotional representations in data physicalizations. By employing countable and uncountable shape parameters, and grouping them into surface texture and overall shape categories, we aim to provide a flexible, expressive system for users to explore how emotional attributes can be translated into tangible 3D forms.

\subsection{Interaction Flow}

To explore the opportunities and challenges of leveraging AI in personal affective physicalization, we designed two versions of PhEmotion. The AI-assisted version incorporates all four design components, DC1, DC2, DC3, and DC4, while, the manual version includes DC4 and partial DC2-3 without AI support.
Below, we present the interaction flows of both versions.

\subsubsection{AI-assisted Version}

Users first engage with a chat interface, where they narrate their emotional experiences and feelings. The system facilitates this process by posing reflective prompts and dynamically responding to user input (DC1). 
After the conversation ends, the system leverages the LLM’s capabilities to analyze the narrative and extract emotional keywords paired with intensity values, based on both explicit and implicit cues in the text (DC2). The model identifies key emotional concepts from the dialogue and uses these to suggest emotions that might reflect the user’s experience. Users can review these AI-suggested keywords, add their own emotional keywords if desired, and then score each emotion on a scale from 0 to 4.5 to indicate intensity (DC3).
With the finalized emotional parametrics, users proceed to the mapping stage, where they connect each emotion to selected shape parameters (DC4). A real-time 3D preview supports iterative refinement before exporting the design for 3D printing.

\subsubsection{Manual Version}
In the manual version of PhEmotion, users take full control of the process. They could input emotional keywords and assign corresponding intensity values without AI support (partial DC2 and DC3). They could also
have their own plan in their mind to directly map their emotions to shape parameters by modifing the slidebar based on their own interpretation (DC4).

Throughout process of both versions, users can iteratively modify keywords, adjust values, and refine their mappings as they develop their design.

\subsection{Implementation}~\label{implementation}
The front-end of PhEmotion is based on the React framework, the Grasshopper® node-based editor\footnote{\url{https://www.grasshopper3d.com}} to generate subdivided surface geometries, and Three.js for real-time 3D rendering, with shape attributes and code inspired from an existing NFT website \footnote{\url{https://blobmixer.14islands.com/}}. 
Models can then be exported in OBJ format to be printed out into artifacts like jewelry or decoration lamps (see Figure \ref{interaction} <c>).

On the back-end, Flask connects the system to GPT-4o\footnote{\url{https://chatgpt.com/}}, which operates in a zero-shot manner, facilitating dynamic and contextually appropriate responses. 
Specifically, the chat agent (DC1) is guided by instructions that prompt emotional exploration, encouraging users to share their experiences and feelings. 
As an emotion physicalization helper, the system responds empathetically, engaging in a natural conversation by acknowledging emotions and following each response with a relevant question, helping users explore their emotions while keeping the conversation focused on their personal experiences, without using markdown or unnecessary repetition. 
The emotional keywords generator (DC2) use GPT-4o to analyze chat logs and automatically generate a list of recommended emotions based on the conversation. 
The system parses user input and outputs a list of 4--7 emotion words in JSON format. 
Prompts make sure the affective intensity analysis (DC3) performs sentiment analysis to quantify the emotional content within user input.
By analyzing the user-generated text, the system outputs a numerical representation of the intensity (from 0 to 4.5, accurate to 0.1) of each identified emotion.

It is worth clarifying that while Figure~\ref{interaction}<c> illustrates a physical 3D-printed artifact, this particular print does not represent an actual participant-generated output from our study. Since fabricating physical artifacts was not our primary study focus, we printed several representative samples to provide participants with tangible reference objects during the study. This decision aligns with our assumption that potential future users of PhEmotion would benefit from interacting with existing physical examples, enabling them to better resonate with and more intuitively map emotions to physical shapes.

For artifact fabrication, we employed white and transparent resin materials for 3D printing. After printing, each piece underwent sanding, polishing, and assembly with metal fittings. The main fabrication constraints arise from the requirement for high-precision finishing, but this workflow is already well-supported by mature and accessible industrial manufacturing services.
    \begin{figure*}[t!]
    \centering
    \includegraphics[width=\linewidth]{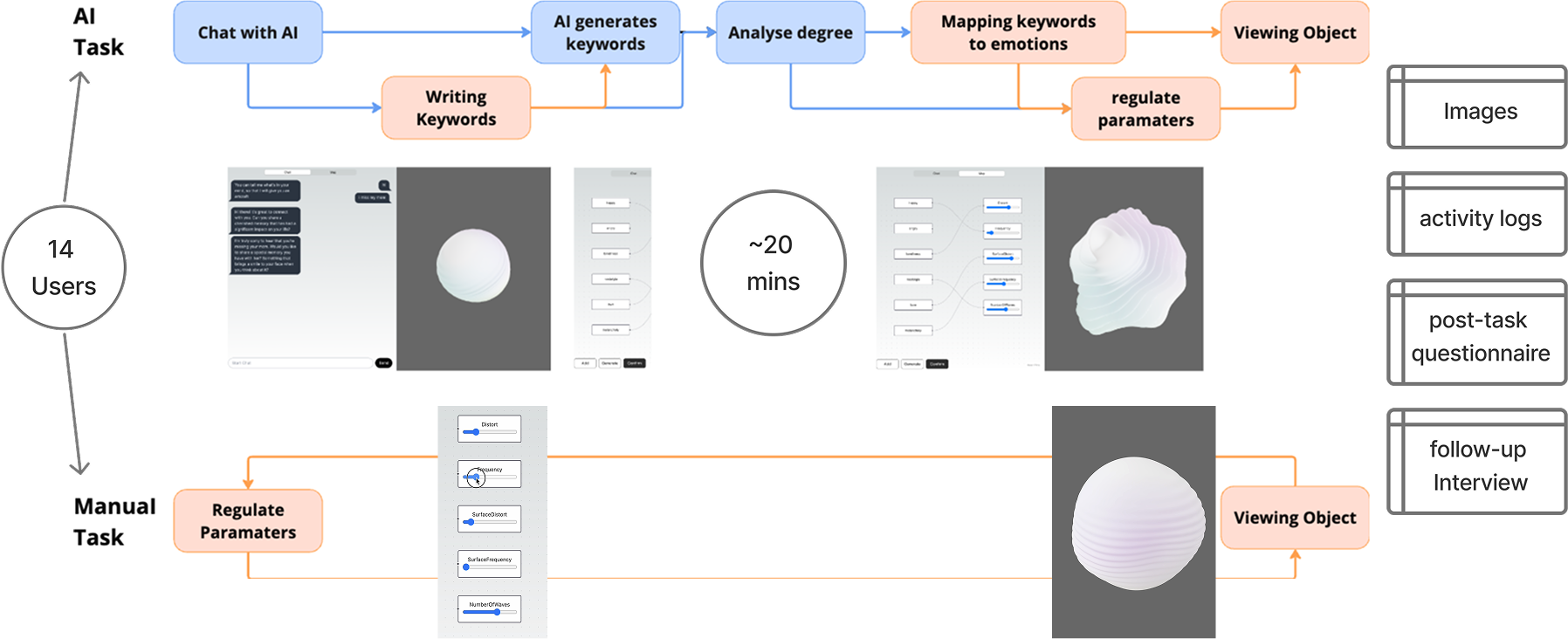}
    \caption{Empirical study on usage flows in PhEmotion with and without AI assistance. In the AI-assisted condition, the AI generates keywords, narrates, and analyzes emotions (blue boxes), while users write keywords, map them to shape parameters, and adjust the final object (pink boxes). In the manual condition, participants create and refine a 3D object by directly manipulating shape parameters, without AI support (all steps in pink).}
    \label{fig:enter-label}
\end{figure*}

\section{Study Methods}

We designed a within-subject comparative study with qualitative measures to uncover how people relate to and engage with a novel, emotionally expressive design tool. 
RtD \cite{zimmerman2007research} approach positions the system as a boundary object: a speculative artifact meant to elicit new perspectives, reflections, and strategies.
Rather than simply asserting that AI-enhanced design ``works'' in a definitive way, the aim was to explore the richness and variety of participant responses.

\subsection{Participants \& Recruitment}

Participants were recruited through social media platforms, where a poster outlining the study’s objectives and activities was shared. The poster called for individuals interested in exploring how AI can assist in the emotional physicalization process and those who desired to encode their memories, thoughts, or emotions into tangible forms. Those who expressed interest in the study contacted us directly to participate. As compensation for their participation, each participant received a \$15 coupon upon completion of the study. 

The target participants were individuals with a need for emotional expression, self-reflection, or interest in creative craft processes, as such people would be typical users of a system like PhEmotion. 
We specifically sought in-person participation to ensure an engaging and interactive experience for each individual.

14 participants (P1--P14) participated in the study. There were 8 female and 6 male, ranging in age from 20 to 34 years. 
Their occupations varied, including fields such as Computer Science, Media, Law, Freelancing, Management, and Civil Engineering. 
Six participants had previously considered methods for recording emotions, 11 had hands-on experience in creative activities like crafting and drawing, and six were familiar with digital manufacturing or modeling software.

\subsection{Study Procedure}
Participants first completed the pre-study questionnaire.
Then, a researcher guided them through 1) a tutorial session, 2) task session I, 3) post-task I questionnaire and interview, 4) Task session II, and 5) post-task II questionnaire and interview. 

In 1) the tutorial session, participants were first introduced to the concept of data physicalization with images from Dataphys Wiki\footnote{\url{https://dataphys.org/wiki/}}. Following this, they were introduced to the study and the PhEmotion system. The session began with an overview of the system interfaces (see Figure \ref{interaction}), which included chat <a> and flow <b>. A demonstration was provided to ensure that participants understood how to use the system effectively. Additionally, examples of 3D printed physical objects <c> were presented to offer a visual and tactile context. Following this, participants were asked to recall two emotionally significant experiences of similar intensity. For example, one might be reflecting on the feeling of joy from a long-awaited reunion with a friend, while another might be recalling the sadness of a difficult farewell. They would then use those two experiences in the task I and task II. This ensured a meaningful basis for comparison. After the tutoral, we conduct pre-task interview and questionnaire to collect general demographic information, such as age, gender, relevant experience, motivation. The goals of the upcoming tasks were discussed, with particular attention paid to the emotional content involved. Finally, participants signed the consent form, formally agreeing to take part in the study.

In 2) task session I and 4) task session II, participants interacted with PhEmotion to design a physicalization. To counterbalance for order effect, half the participants completed the task with AI assistance first and then the manual condition, while the other half adopted the reverse order. 

In the AI-assisted condition, participants had full access to the system, including the chat interface, emotional keyword extraction, and sentiment analysis intensity values. In the manual condition, AI features were disabled: participants had to come up with emotional dimensions and construct mappings with shape parameters manually. Those two task sessions concluded when participants expressed they were satisfied with their designs and lasted 10--20 minutes each.

In 3) post-task I questionnaire and interview and 5) post-task II questionnaire and interview, participants completed a brief questionnaire and participated in a short follow-up interview. 
The Likert-scale questionnaire asked participants to rate the following questions on a scale of 1 (for strongly disagree) to 7 (for strongly agree): 1) Ease of Design: Was it easy to design the shape? 2) Emotional Awareness: Did the task help you clearly identify the emotions you were expressing? 3) Shape Satisfaction: Are you satisfied with the form you created? 4) Emotional Representation: Does the shape adequately represent your emotions?
The post-task interviews focused on the creative process, emotional resonance, and interpretation of the resulting object, as well as their perceptions of accuracy, building on participants' responses to the questionnaire. 

At the end of the study, a final semi-structured interview was conducted to synthesize insights and contrast conditions.

\subsection{Data collection and analysis}

PhEmotion does not automatically store participants’ emotional narratives or conversation logs. Participants explicitly indicated their privacy preferences through a consent form prior to the study, deciding whether or not to allow screen recording of their sessions, including chat content and interactions. They had the option to pause or terminate the recording at any point during the session, or to request recording of interface interactions only, excluding the conversational interface.

The qualitative data we collected consists of (1) transcripts from 8.6 hours of pre-, post-task, and final interviews, and (2) transcripts of think-aloud data and observations from 9 hours of screen recordings of the task sessions.

We conducted a thematic analysis of the think-aloud and observation data and of the post-task interviews using a bottom-up, inductive approach, following Braun and Clarke’s six-phase framework for thematic analysis~\cite{clarke2017thematic}.

The analysis followed a qualitative approach, focusing on the interpretation of meaning rather than the frequency of specific codes.  
We first create a physicalization design summary based on the input and output of 14 users (see Table~\ref{tab:participant_data}) to better interpret and associate it with their interview content.
To ensure credibility, the coders employed triangulation across multiple data sources and engaged in reciprocal review of the coding themes and interpretations.
We prioritized understanding the diverse ways participants interacted with the system, interpreting their behaviors and language to extract emergent themes related to emotional mapping, automation, and authorship.

First, two coders independently reviewed the data from four participants (P1, P6, P9, and P13), which, based on initial observations, were good representative of all participants, based on initial observations. 
Each coder carefully watched the corresponding screen recordings, took analytic field notes, and read the transcripts multiple times to gain an in-depth understanding. 
The two coders then independently applied open coding to generate descriptive codes that can capture the nuances of the data.

Then, the two coders met with a third researcher to compare and discuss the initial codes. 
Through iterative discussion and consensus-building, they developed a first set of codes. Examples of generated codes include ``\textit{small ups and downs (frequency attributes) represent negative emotions}'' and ``\textit{Emotions cannot be fully quantified with index}''.

In the next step, the remaining data were split between the two coders. 
Each coded their assigned portion with the references of initial codes. 
They met regularly to compare coding outcomes, refine definitions, and ensure consistency. 
Any uncertainties or disagreements were resolved collaboratively or escalated to the larger research team for discussion. 

In the last step, all authors gathered in multiple meetings to review the generated codes and identify high-level themes. 
Themes were iteratively refined to ensure they accurately represented the data and aligned with the research objectives.

To analyze the results from the post-task questionnaires, we use descriptive statistics rather than inferential statistics, given the relatively low power such analysis would yield.

\begin{table*}[h]
    \centering
    \vspace*{4pt}
    \caption{Example of personal affective physicalization design showing inputs and outputs of participant P1 under both AI-assisted and manual conditions (P2-P14 can be found in the supplementary material). The metrics represent the generated emotional keywords, their corresponding intensity values, and the shape parameters those are mapped to. The ‘Content’ is a brief description of the emotional experiences encoded by the participant under each condition.}
    \label{tab:participant_data}
    \begin{tabular*}{\textwidth}{@{}m{0.04\textwidth} m{0.10\textwidth} m{0.29\textwidth} m{0.145\textwidth} m{0.16\textwidth} m{0.13\textwidth}@{}}
        \toprule
        \textbf{ID} & \textbf{Content for AI Task} & \textbf{AI-generated metrics} & \textbf{Screenshot of AI-assisted output} & \textbf{Content for Manual Task} & \textbf{Screenshot of Manual output} \\
        \colrule
        P1 & An interesting and nostalgic dream & Nostalgia(4) - NumberofWaves; \newline Happiness(3.5) - SurfaceFrequency; \newline Anticipation(3) - Frequency; \newline Worry(2) - Distort; \newline Satisfaction(3) - SurfaceDistort & \includegraphics[width=0.145\textwidth]{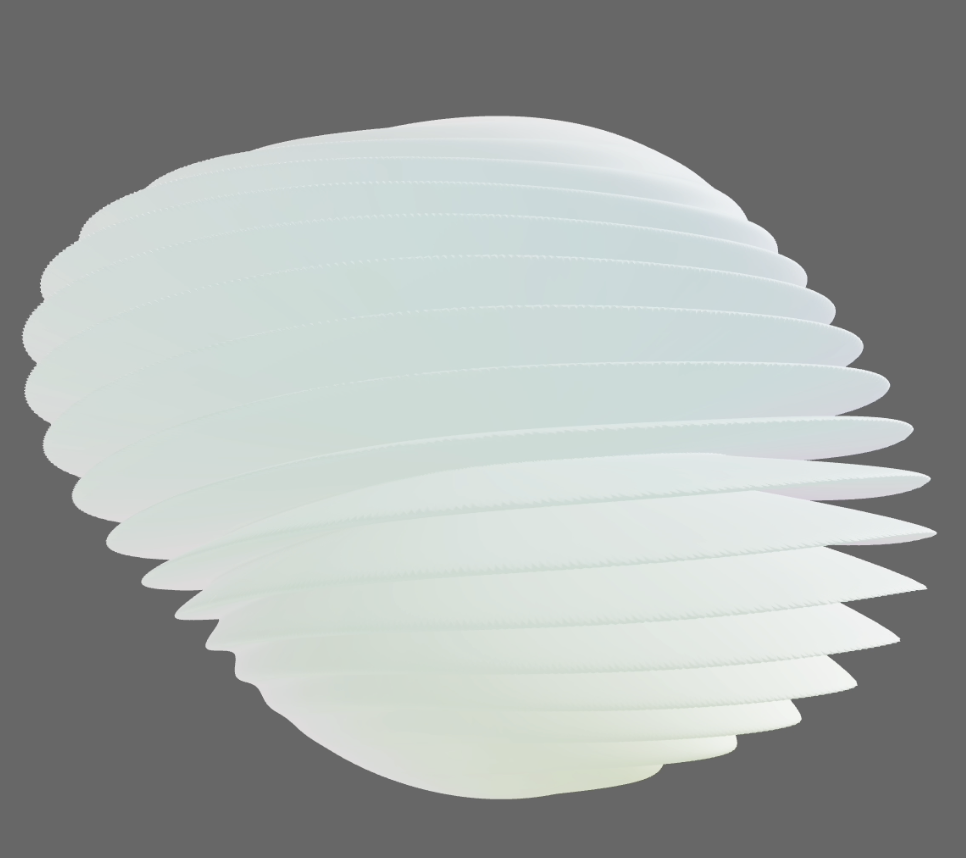} & A nightmare ... feeling an intensifying sense of disgust and fear beneath a composed exterior. & \includegraphics[width=0.13\textwidth]{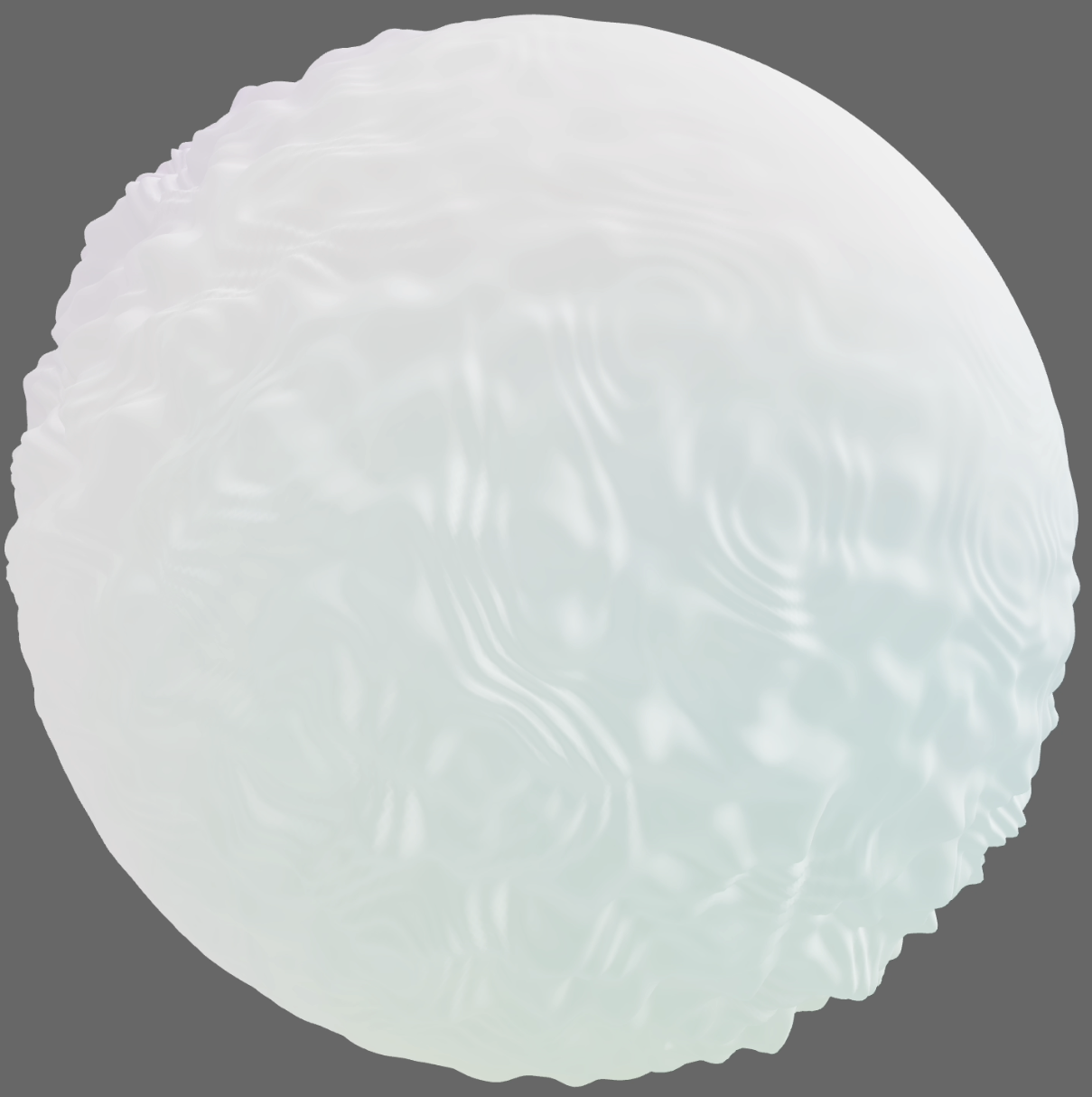} \\
        \botrule
    \end{tabular*}
    \vspace*{8pt}
\end{table*}

    \section{Results}

\begin{figure}
    \centering
    \includegraphics[width=\linewidth]{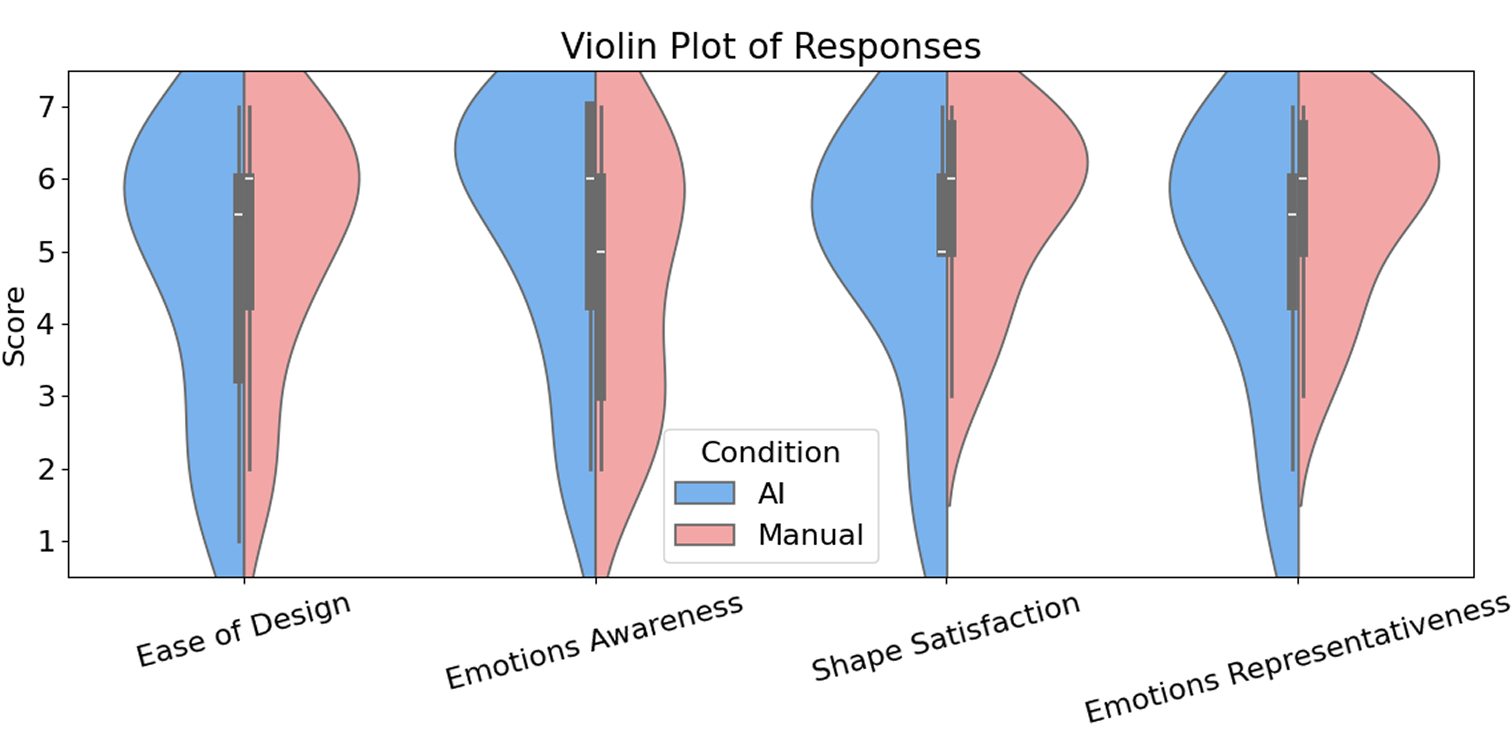}
    \caption{Violin charts of questionnaire responses for four questions and for the two conditions (AI-assisted and manual).}
    \label{fig:chart}
\end{figure}

From the Likert scale responses (summarized in Figure~\ref{fig:chart}), we can see that participants gave slightly higher scores for emotion awareness in the AI-assisted condition than in the manual condition. On the other hand, they more satisfied with the final shape and with the representation of their emotions with the manual condition. These descriptive quantitative results are helpful in contextualize the qualitative findings. 

\subsection{Creative Mapping and the Gap Between Language and Form}

The challenge of translating emotions into shapes and forms was a recurring theme in the participants’ creative processes. Many participants expressed satisfaction with their ability to map emotions onto shapes, although the process often involved navigating a gap between abstract emotional language and concrete visual representations.

\subsubsection{Translating Emotions into Objects}
A prominent factor to this gap is intuitive versus semantic (low-level) mapping. 
In the manual condition, participants often relied on \textbf{intuition}, exploring the shapes as they emerge. P1 described this process as essentially ``blindfolded,'' where they adjusted parameters based on immediate visual impressions rather than a calculated strategy. 
P6 highlighted the spontaneity of their manual approach, adjusting shapes based on first impressions: \textit{``If I see it and feel I don’t want it to be printed, then I’ll adjust it.''}

In contrast, in the AI-assisted condition, participants found themselves more driven by \textbf{semantic mapping}, in which the AI-generated emotional keywords and intensity values strongly influenced their decisions. 
P3, for instance, felt that the AI’s keyword analysis closely matched their emotions, saying: \textit{``I tried it out and found that the analysis was somewhat similar to how I feel.''} Other participants linked specific emotional keywords to their shape decisions. For example, P12 said \textit{``Serenity'' [AI-generated keyword] encompasses multiple emotions, this word encapsulates everything I aim to convey.''} This shows how the AI’s structured approach helped participants summarize language (narrative) to dimensions more directly, along with the results still open to personal interpretation.

Another key aspect of mapping emotions to physical properties was sensation-driven or \textbf{synesthetic mapping}, where emotions like anxiety were linked to specific visual qualities. 
P9 associated \textit{``blunt edges''} with warmth and safety, contrasting with \textit{``round shapes''}, which felt insecure and slippery. 
Similarly, P13 linked a painful memory to a sharp, piercing shape, mirroring the intensity of their emotional experience, saying ``\textit{It’s like the feeling of being pierced by a sea urchin.}''

\subsubsection{Mapping Layered Emotions to Shape Parameters}
Several participants made a clear distinction between \textbf{primary and secondary emotions}, and when they did, they considered primary emotions more important, and paid more attention to those in the design process. For example, P8 double-encoded ``love'' as they thought it was the most important emotion, saying \textit{``I connected `love' to two parameters because I felt it was the most fundamental emotion, so I wanted to use it more.''}
P14 grouped emotions based on their relationships, combining ``nostalgic'' and ``touched'' as more primary emotions. 
In terms of visibility, P4 made ``love'' the most prominent emotion, saying \textit{``I feel that love must be in the first place and should be directly seen from the surface. That’s why I chose to map it to `surface distort'.''} 
P3 also linked surface distortions to primary emotions, saying \textit{``I felt it represented a primary part of my emotion.''} In all those cases, participants wanted to tie primary emotions to prominent visual elements in their designs.

Participants also categorized emotions as either \textbf{internal or external}. 
P6, for instance, associated emotions like ``confusion'' with internal shape changes, while emotions such as ``hope'' were reflected on the surface, saying: \textit{``` lost' and `confused' are something inner to me, but I am eager for change. So I would like to correspond them to the two attributes of this overall shape, respectively.} And for `hope`, `passion', and `desire', I would like to correspond them to the surface attributes. 
P5 built a base shape reflecting security, then added external emotions through surface textures. They used a ``bridge'' notion to connect these internal and external layers, saying: \textit{``After talking with him, he gave me these words… I started with `tolerance,' because it connects the positive and negative words. I felt `tolerance' was the bridge between them, so I placed it in the center. The positive words, `security' and `warmth,` set the tone for the shape... The other words relate to the surface textures, interwoven with the emotions. `Tolerance' acts as a bridge, connecting everything.''}

Collectively, participants showed a nuanced understanding of emotions, with some focusing on primary emotions and others exploring the internal-external divide.

\subsubsection{Mapping Emotions is Challenging} 
While participants did successfully map emotions to physical parameters, this was not a straightforward process and many participants expressed \textbf{difficulty in translating emotions into shapes}. 
For example, P4 said, \textit{``I didn’t know how to link the emotions expressed in words to their visual counterparts.''} 
Others, like P6, found the mapping process confusing, especially when trying to connect the AI-provided emotion metrics and the shape parameters. 
Several participants also said they wanted to express more abstract emotions like ``loneliness'' or ``confusion'', noting that the available shape parameters did not fully capture these complex emotional states. For example, P12 said \textit{``I don’t know how to use these parameters to express `loneliness'''.}

This mapping gap was often bridged with metaphors and symbolic interpretations. 
For example, P7 described their design as a \textit{``shell with a pearl''} that symbolized their relationship with their grandparents. 
Similarly, P3 likened their emotional experience to the way \textit{``rain leaves tiny impressions on soft surfaces''}, reflecting the subtle and cumulative nature of their mood. These metaphorical interpretations allowed participants to overcome the limitations of the mapping system by using abstract visual cues to represent deeper, more complex emotional states.

\subsection{Creative Autonomy vs. AI Guidance}
A noticeable tension emerged between trusting the AI to generate outputs and maintaining full control over the creative process. This conflict highlights the broader challenge of integrating automation while preserving authorship, particularly in tasks that involve personal emotions and expressions.

Several participants appreciated the AI’s suggestions as a starting point, noting that it \textbf{helped them bypass some of the more labor-intensive parts} of the process. 
P1 explained \textit{``I felt confident about the keywords, so I skipped deeper thinking processes.''} P6 also said \textit{``the overall analysis from AI aligns well with my own perceptions,''} and P9 said \textit{``I relied on them directly and initially felt that the extracted words had a degree of trust, even though I later realized they were somewhat alike.''}
This sentiment echoes a general theme where AI’s structured approach reduces the need for participants to engage in more complex decision-making.

However, others voiced concerns about the \textbf{AI’s influence on their creative autonomy}. As P5 said, "Without AI, it’s freer; there are no limitations" regarding how to express and convey emotions.
Similarly, P8 explained that manual process felt more personal and fulfilling when it unfolded gradually through their own efforts, saying \textit{``AI gives me a numerical value, and then I create from that; it’s more immediate, but also less personal.''}. And P9 said \textit{``I prefer my own (manual) version,''} suggesting a stronger attachment to self-authorship.
This highlights the value some participants placed on the organic nature of their creative process, where decisions and changes are made intuitively rather than algorithmically.

This reflects a broader struggle over authorship. Several participants were uncomfortable with AI’s influence on their emotional expression, feeling it changed their original intentions. 
For example, P1 explained \textit{``Chatting with AI shifted my original idea,''} and P4 said \textit{``I didn’t want to be told how to feel.''}

On the positive side, some participants found the AI-assisted condition to be particularly useful when they were \textbf{seeking clarity}. 
The AI’s structured breakdown of emotions into smaller, manageable components helped some participants articulate their feelings more effectively. 
For example, P3 said \textit{``With AI, it’s like it keeps asking you for details, where you were, what you saw, what you heard, organizing emotions layer by layer.''}

\subsection{Metrics vs. Meaning}

The third tension our results highlight is the tension between emotional awareness and the reduction of complex feelings into quantifiable data. 
Although the AI-generated metrics with emotion keywords and numeric values did offer participants a way to enhance their understanding of their emotions, participants also questioned the meaning of those.

As P5 noted, the numeric data offered a clearer understanding of their emotional landscape: \textit{``When I was manually adjusting, I felt it was too sharp, and it made me uncomfortable, so I denied it. For example, while the chat did measure a high degree of negative emotion, if I were adjusting it manually, I would intentionally soften that emotion.''} This suggests that AI’s ability to measure emotions brought a level of self-awareness, but also provoked a more conscious effort to regulate or modify emotional responses.

However, some participants were skeptical about the idea of reducing emotions to numerical representations, and concerned that an \textbf{AI data-driven approach strips away the nuances and complexity inherent in human emotional experiences.}
For example, P13 said \textit{``A person is a creature full of emotions. I think it is impossible to be completely objective, like AI.''} 
P1 echoed this concern, stating, \textit{``The process of disassembling emotions feels unnatural, like trying to create a word for the fibers on a banana—it doesn’t capture the meaningful nuance.''}

Additionally, participants questioned the validity of the AI’s emotion analysis, with some finding \textbf{the numeric outputs ``incorrect'' or overly simplified.} 
As P10 described, \textit{``keywords of sentiment analysis are sometimes inaccurate and there is a situation of over-analysis.''} 
P13 further elaborated, \textit{``It can’t be simplified into a single number,''} suggesting that the depth and layers of human emotions could not be adequately represented by simple metrics.

Despite these critiques, many participants still found value in AI’s ability to generate data points, even if imperfect. Some participants, like P12, admitted that breaking down emotions into structured forms was challenging; they said: \textit{``I’m not very good at breaking down emotions into structures; it’s not how I naturally approach things.''}

    \section{Discussion}
Rather than demonstrating the superiority of AI in supporting in personal affective physicalization design, our intention was to create a tool as a provocation to stimulate deeper discussion around our three central questions raised in the introduction:
a) How do people create a physical embodiment of complex emotions?,
b) Does Human-AI co-creation benefit personal affective physicalization? and
c) Is there an interplay between automation vs. manual control

\subsection{Physical Embodiment of Complex Emotion}
We gained insights into how participants map emotion into physical forms with and without AI. AI assistance encouraged them to adopt a more semantic-centric approach where participants would first use words to describe the feelings. While the manual condition encouraged a more intuitive making process, where participants directly manipulated the shapes until the shape matched their feeling. This intuitive process allowed participants to perceive a direct association between the artifact’s shape and their emotional states.

While early work in data physicalization used simple shapes such as bar charts~\cite{djavaherpour2021}, our layered shape design involving both surface and body parameters uniquely elicited layered emotional expressions. 
This indicates the importance of the design medium for personal affective physicalization and reveals potential opportunities for enriching the shape vocabulary, such as discrete features like wave count and continuous variables like surface distortion, can expand the expressive possibilities of personal affective physicalization.
While parametric design introduces a higher barrier to creation, tool support like ours could mitigate part of the challenge.

\subsection{Human-AI Co-creation in Personal Affective Physicalization}

We observed human-AI co-creation unfolding in a unique way when it entered this personal and intimate space of creation and meaning-making. When humans are unclear or have difficulty articulating their emotions, AI could help them gain clarity. When there is no straightforward mapping between semantic emotion and shape parameters, participants made use of metaphors and symbolic interpretations to fill the gap, where creativity emerges and human meaning-making or narrative-creation takes place.

\subsection{Automation vs. Manual Control}

We found that participants valued both automation and manual control, although not uniformly. It was appreciated that automation gave participants a starting point and reduced labor-intensive parts. Manual creation enhanced ownership and personal connection with the artifact. The design implication of this for future AI support systems entering this context is to strike a balance between computational efficiency and spaces for interpretive ambiguity, preserving the expressive freedoms that are central to personal and affective design. Completing the artifact is not the primary goal of personal affective physicalization. Rather, it is the process of creation, emotional engagement, and reflection that gives the experience its value.

\subsection{Challenges of Co-Creation}

This research is in response to the larger trend of increasingly embedding AI in creative and expressive domains, with acknowledged risks and ethical implications.
Our main goal in this research was to ask: what do people actually experience when AI enters the terrain of personal affective physicalization design?

Our results make it clear that there is no straightforward answer to that question. 
Participants did not reject AI outright: many found value in its suggestions. 
However, they also expressed discomfort when automation interfered with emotional intuition or disrupted their creative flow. 
For them, emotional expression was not merely about producing a form, but about following a feeling. AI, as a system built on structured pipelines, can struggle to accommodate the nonlinear, intuitive nature of that process.

\subsection{Toward Emotionally Meaningful Human-AI Collaboration}
The central insight of this work is that emotion, when represented computationally, should not be treated solely as input to be optimized. It is also a lens, a material, a narrative. As has been discussed in prior work~\cite{park2025reimagining}, people reflected on the context and emotions associated with their data through the AI-generated images based on their personal data. And people use various approaches, including abstract, symbol, and scene, to represent emotion \cite{lee2017designing}. 
Our participants engaged with emotional data not to reach a correct solution, but to make sense of themselves.

\subsection{Limitations}

This study is exploratory in nature. The LLM’s integration into the workflow is early-stage, and user familiarity with both AI and parametric modeling varied. Our participants were visualization novices, and while this made their perspectives especially revealing, it also limits generalizability.

Another limitation of our work is the nature of a lab study lacking longer-term engagement with the participants. We concentrated on the design aspect of personal affective physicalization, showcasing a 3D view of the display on screen. However, this approach did not include the actual physical construction of the object. Moving forward, it would be beneficial to explore how individuals feel when they can interact with the objects they create. Understanding these emotional responses could provide valuable insights into how effectively the physicalizations convey their intended emotions. Future studies will let them print out their created artifacts, support a few iterations and collect data about how they made use of the objects.

While our study explores where automation may support personal affective physicalization, it does not prescribe which steps should be automated. Our aim was to investigate how people engage with automation in emotionally expressive design, rather than to build a fully automated system. To this end, we intentionally restricted AI assistance to the design phase while leaving the data construction phase fully manual. Although this choice was motivated by the desire to preserve user agency, our findings suggest that it may have introduced a sense of disjunction for some participants. Several participants expressed uncertainty about how to reconcile AI-generated numerical suggestions with their own manual adjustments, perceiving semi-automation and full manual control as two divergent modes of thinking. Notably, one participant (P6) mentioned that they would have liked to see a fully AI-generated sphere, describing it as akin to a ``fortune-telling'' experience. These findings suggest the importance of carefully balancing automation and authorship in future systems, where partial automation may risk confusing or alienating users.

\subsection{Future Work}
Our aim is not to declare whether AI-assisted emotional design is good or bad. Rather, we want to capture emerging attitudes and strategies as people encounter this new hybrid space. Importantly, the ``failure'' of AI to feel emotionally resonant for some participants is not a negative result—--it is a clue. It suggests that design systems for emotional expression must be less prescriptive and more dialogic, allowing space for contradiction, confusion, and personal narrative.

This points to a new research direction: designing for emotionally rich, personally meaningful co-creation between humans and AI—without reducing either to a tool. Promising areas of future work we have identified include:
\begin{itemize}
    \item Examining which aspects of emotion physicalization benefit from automation, and which should remain manual to preserve interpretive depth.
    \item Supporting novice users in learning parametric modeling for expressive purposes, especially when transitioning from tactile or analog mediums.
    \item Understanding how people intuitively map affective states to shape parameters and spatial metaphors in digital and physical environments.
\end{itemize}

\section{Conclusion}
Through the design and implementation of PhEmotion, an LLM-assisted personal affective physicalization system, we contribute a novel AI-assisted workflow that supports non-expert users in creating emotion-driven physical representations. Our qualitative study reveals how individuals navigate the tensions in personal affective physicalization design. These tensions are: the choice between automation and manual control, the gap between emotions, language and the shape representation, and the tension between emotional awareness and the reduction of emotions into quantifiable form.   

Our findings show that AI assistance can reduce technical barriers and support clarity in emotion reflection, while the fundamental work of meaning-making remains with human creators. Rather than replacing human expression, AI integration reshapes how users approach affective design, transforming emotion data into materials for metaphor and self-reflection. This research contributes to understanding how AI can support creative processes in ambiguous design spaces while preserving the inherently personal nature of emotional expression and narrative construction.

\section{Acknowledgement}
This research was funded in part by NNSFC-Young Scientists Fund (CityU 62202397), NFRFR-2022-00570 (A Co-Design Exploration), NSERC Discovery Grant: Interactive Visualization RGPIN-2019-07192, and Canada Research Chair in Data Visualization CRC-2019-00368.  



    \vspace*{-8pt}

    \begin{IEEEbiography}
        {Ruishan Wu}{\,}is a PhD student in the Department of Computing Science, Simon Fraser University, Canada. She is interested in human computer interaction, artificial intelligent. Her expertise spans digital visuals, hardware manufacturing, and user experience, with a research focus on interactive visualization and data physicalization.
        Contact her at ruishan\_wu@sfu.ca

    \end{IEEEbiography}

    \begin{IEEEbiography}
        {Zhuoyang Li}{\,} is a PhD student in the Department of Industrial Design at Eindhoven University of Technology (TU/e), the Netherlands. Her research focuses on how to design personalized healthcare technology to empower people’s well-being in everyday life, and how these technologies can adapt to users’ changing needs and personal contexts. Contact her via z.li7@tue.nl.
    \end{IEEEbiography}
    
    \begin{IEEEbiography}
        {Charles Perin}{\,} is an Associate Professor with the University of Victoria, Canada. 
        His research focuses on designing and studying new interactions for visualizations and on understanding how people may make use of and interact with visualizations in their everyday lives, including mobile and physical visualization.
        Contact him at cperin@uvic.ca.
    \end{IEEEbiography}

    \begin{IEEEbiography}
        {Sheelagh Carpendale}{\,} holds a Canada Research Chair: Data Visualization at Simon Fraser University. Her research in data visualization, data empowerment, collaborative visualization, and large interactive displays draws on her dual background in computer science and the visual arts. Contact her at sheelagh@sfu.ca.
    \end{IEEEbiography}

    \begin{IEEEbiography}
        {Can Liu} {\,} is an Associate Professor at the School of Creative Media, City University of Hong Kong. Her research focuses on the design and empirical study of future interfaces involving multimodal input and unconventional displays. She is the corresponding author of this article. Contact her at canliu@cityu.edu.hk.
    \end{IEEEbiography}\vfill\pagebreak
\end{document}